%% file: TD-LM-PAM_ArXiv.tex
\documentclass[conference]{IEEEtran}
\IEEEoverridecommandlockouts

\usepackage{cite}
\usepackage[utf8]{inputenc}
\usepackage{amsmath,amssymb,amsfonts}
\usepackage{algorithmicx}
\usepackage{algpseudocode}
\usepackage{graphicx}
\usepackage{textcomp}
\usepackage{xcolor}
\usepackage{tikz}
\usepackage{graphicx}
\usepackage{caption}
\usepackage{tabularx}
\usepackage{adjustbox}
\usepackage{booktabs} 
\usepackage{makecell}
\usepackage{array}
\usepackage{soul}
\usepackage{dblfloatfix}

\def\BibTeX{{\rm B\kern-.05em{\sc i\kern-.025em b}\kern-.08em
 T\kern-.1667em\lower.7ex\hbox{E}\kern-.125emX}}

\usepackage[colorlinks=true, linkcolor=blue, citecolor=blue, urlcolor=blue]{hyperref}

\definecolor{PastelBlue}{RGB}{173, 216, 230} 
\definecolor{PastelPink}{RGB}{255, 182, 193} % Light pastel pink
\definecolor{RoyalBlue}{RGB}{30,77,216}
\definecolor{SunsetOrange}{RGB}{255,107,26}
\definecolor{TurquoiseGreen}{RGB}{0,194,160}
\definecolor{GoldenYellow}{RGB}{255,217,61}
\definecolor{FuchsiaPink}{RGB}{230,57,143}

\newcolumntype{Y}{>{\centering\arraybackslash}X}

\begin{document}

\title{Time-Domain Linear Model-based Framework for Passive Acoustic Mapping of Cavitation Activity}
\author{Tatiana Gelvez-Barrera,
Barbara Nicolas, Denis Kouamé,
Bruno Gilles, Adrian Basarab
\thanks{This work was supported by the LABEX CELYA (ANR-10- LABX-0060) and LABEX PRIMES (ANR-11-LABX-0063) of Université de Lyon, within the program ``Investissements d'Avenir" (ANR-11-IDEX-0007) operated by the French National Research Agency (ANR), as well as the ``CAVIIAR" Project (ANR-22-CE19-0006), operated by the French National Research Agency (ANR).}
\thanks{T. Gelvez-Barrera is with IRIT, Université de Toulouse, CNRS, Toulouse, France, and Université Claude Bernard Lyon 1, INSA Lyon, CNRS, Inserm, CREATIS UMR 5220, U1294, F-69100, Lyon, France.
(e-mail: tatiana.gelvez@creatis.insa-lyon.fr).}
\thanks{B. Nicolas and A. Basarab are with Université Claude Bernard Lyon 1, INSA-Lyon, CNRS, Inserm, CREATIS UMR5220, U1294,
F-69100, Lyon, France. (e-mail: barbara.nicolas@creatis.insa-lyon.fr, adrian.basarab@creatis.insa-lyon.fr). }
\thanks{D. Koumé is with IRIT, Université de Toulouse, CNRS, Toulouse, France. (e-mail: denis.kouame@irit.fr).}
\thanks{B. Gilles is with LabTau, Inserm, U1032,
Université Claude Bernard Lyon 1, F-69003, Lyon, France. (e-mail: bruno.gilles@inserm.fr).}
}

\maketitle

\begin{abstract}
Passive acoustic mapping enables the spatial mapping and temporal monitoring of cavitation activity, playing a crucial role in therapeutic ultrasound applications. Most conventional beamforming methods, whether implemented in the time or frequency domains, suffer from limited axial resolution due to the absence of a reference emission onset time. While frequency-domain methods, the most efficient of which are based on the cross-spectral matrix, require long signals for accurate estimation, time-domain methods typically achieve lower spatial resolution. To address these limitations, we propose a linear model-based beamforming framework fully formulated in the time domain. The linear forward model relates a discretized spatiotemporal distribution of cavitation activity to the temporal signals recorded by a probe, explicitly accounting for time-of-flight delays dictated by the acquisition geometry. This model is then inverted using regularization techniques that exploit prior knowledge of cavitation activity in both spatial and temporal domains. Experimental results show that the proposed framework achieves enhanced or competitive cavitation map quality while using only 20\% of the data typically required by frequency-domain methods. This highlights the substantial gain in data efficiency and the flexibility of our spatiotemporal regularization to adapt to diverse passive cavitation scenarios, outperforming state-of-the-art techniques.
\end{abstract}


\section{Introduction}

Cavitation activity refers to the acoustic emissions produced by gas or vapor microbubbles within a liquid medium~\cite{denner2023modeling}. The microbubbles form in biological tissues or originate \textit{in situ} as a result of rapid pressure variations induced by an ultrasound field~\cite{gharat2022microbubbles}. Depending on acoustic pressure and physical properties of the medium, microbubbles may either oscillate non-destructively over multiple cycles, known as non-inertial cavitation, or collapse violently within a few microseconds, known as inertial cavitation~\cite{tan2024modelling, coussios2008applications}. Each regime connotes distinct nature, which is closely related to therapeutic outcomes, bio-effects, or pathological conditions~\cite{saletesresearch}.

Passive Acoustic Mapping (PAM), depicted in Fig.~\ref{fig:PAM} maps the cavitation activity by beamforming radio-frequency (RF) signals passively recorded by a probe. This technique has proven effective for monitoring and guiding ultrasound-based therapies that induce cavitation, such as high-intensity focused ultrasound and ultrasound-enhanced drug delivery~\cite{moonen2025focused}.

\input{figures/figure1}

Ultrasound beamforming techniques can be broadly categorized into time-domain (TD) and frequency-domain (FD) approaches, depending on the domain in which RF signals are processed. TD methods operate directly on raw temporal data, while FD methods analyze signals in the Fourier domain~\cite{lu2018passive, haworth2016quantitative, abadi2018frequency}. These beamforming paradigms have been adapted for cavitation map reconstruction. One of the earliest approaches extended Time-Exposure Acoustics (TEA) to localize cavitation regions, giving rise to TEA-PAM~\cite{gyongy2009passive}. Subsequently, the spatial resolution was enhanced in~\cite{gyongy2011passive} by sparsifying RF signals prior to beamforming. Building on this foundation, the Robust Capon Beamformer (RCB) was adapted in~\cite{coviello2015passive}, introducing RCB-PAM to improve artifact suppression. Later, the Phase Coherence Factor (PCF) was incorporated in~\cite{boulos2018weighting}, developing PCF-PAM to suppress incoherent noise. In parallel,~\cite{crake2018passive} applied Compressive Sensing (CS) theory, proposing CS-PAM that models cavitation emissions as sparse signals for high-resolution reconstruction from few measurements. Complementarily,~\cite{polichetti2020use,sivadon2020pisarenko,du2009review} showed that using the Cross Spectral Matrix (CSM) enables adaptive beamforming in PAM, improving resolution and contrast at low computational cost.

More recently, an increasing number of studies have demonstrated the effectiveness of linear model-based formulation and regularized inversion techniques that formulate beamforming as an inverse problem in related applications, including active ultrasound imaging~\cite{szasz2016beamforming, goudarzi2023unifying} and room acoustics~\cite{gelvez2025time}. In the context of PAM, the most recent contribution adapts the CSM Fitting method, leveraging a FD linear model-based approach to improve spatial resolution~\cite{lachambre2024inverse}.

However, current PAM approaches still face inherent limitations. In particular, achieving high resolution along the axial dimension, perpendicular to the probe, remains challenging due to the absence of a reference emission time. Additionally, while TD methods are well-suited for non-stationary phenomena, they provide lower spatial resolution~\cite{polichetti2018advanced}. In contrast, FD methods, though efficient, are limited by assumptions of stationarity that may not hold in dynamic conditions~\cite{singh2025enhancing}.

Therefore, this work proposes a beamforming framework for PAM, based on a linear forward model fully formulated in the time domain and employing regularized inversion, as illustrated in Fig.~\ref{fig:PAM}.
This framework is hereafter referred to as Time-Domain Linear Model-based Passive Acoustic Mapping (TD-LM-PAM). To this end, we introduce an original linear forward model that incorporates the time-of-flight delays dictated by the acquisition geometry, linking the distribution of cavitation activity to the recorded RF signals.

The proposed TD-LM-PAM framework opens the possibility to incorporate prior knowledge through regularization, effectively capturing spatial and temporal characteristics of acoustic emissions generated by cavitation activity. Accordingly, we investigate several regularization strategies, including sparsity, total-variation, and regularization by denoising.

We perform comparative experiments against state-of-the-art TD and FD methods, showing that our TD-LM-PAM framework achieves enhanced or competitive performance with appropriate regularizers, even while relying on only 20\% of the data required by FD approaches. These results highlight the benefits of incorporating spatiotemporal regularization to improve cavitation map quality, exploiting prior knowledge of cavitation activity and adapting to different scenarios.

The main contributions of this paper include:
\begin{itemize}
 \item An original linear forward model, operating in the time domain, of the acoustic propagation of the cavitation emissions based on the geometry of the acquisition setup, in Section~\ref{subsec:forward}.
 \item A general TD-LM-PAM framework that is flexible and can be adapted to different scenarios through the choice of appropriate spatiotemporal regularization terms, in Section~\ref{subsec:inverse}.
 \item An improved or comparable quality of reconstructed cavitation maps using only 20\% of data typically used for state-of-the-art methods, in Section~\ref{sec:Experiments}.
 \end{itemize}
 
\section{Time Domain Linear Model-based Beamforming Framework for Passive Acoustic Mapping}
\label{sec:proposal}

This section presents the discrete formulation of the proposed TD-LM-PAM framework for estimating cavitation source activity. The framework builds on the proposed linear forward model in Section~\ref{subsec:forward}, and addresses beamforming through regularized inversion as in Section~\ref{subsec:inverse}.

Figure~\ref{fig:notation} schematizes the notation convention used throughout the paper. In what follows, $(x, y, z)$ refer to the lateral, azimuthal, and axial spatial dimensions, respectively, while $(t)$ represents the temporal dimension.

\input{figures/figure2a}

Let $\mathbf{Y} \in \mathbb{R}^{N_m \times N_t}$ be the matrix containing the RF signals recorded by an ultrasound linear array with $N_m$ sensors located at $\vec{\mathbf{r}}_m = (x_m,\, 0,\, 0)$, sampled at $N_t$ temporal samples with a sampling frequency $f_s$, so that, the $k^\text{th}$ temporal sample refers to the time instant $t_k = k / f_s$.
For notational clarity:
\begin{itemize}
 \item $\mathbf{Y}_{:,k} \in \mathbb{R}^{N_m}$ denotes the signal recorded by all sensors at the fixed time instant $t_k$, for $k = 1, \ldots, N_t$.
 \item $\mathbf{Y}_{m,:} \in \mathbb{R}^{N_t}$ denotes the signal recorded by sensor $m$, over all time instants, for $m = 1, \ldots, N_m$.
\end{itemize}
The RF signals can be expressed in vectorized form as:
\begin{equation}
\mathbf{y} \in \mathbb{R}^{N_m N_t} =
\begin{bmatrix}
\mathbf{y}_1^\top & \ldots & \mathbf{y}_m^\top & \ldots & \mathbf{y}_{N_m}^\top
\end{bmatrix}^\top,
\end{equation}
where $\mathbf{y}_m \in \mathbb{R}^{N_t}$ denotes the vector form of the signal $\mathbf{Y}_{m,:}$. 

Let $\boldsymbol{\mathcal{X}} \in \mathbb{R}^{N_x \times N_z \times N_t}$ be the datacube representing cavitation activity in terms of amplitude within a 2D spatial plane evolving in time. The cube is defined along the lateral, axial, and temporal dimensions $(x, z, t)$, where $N_x$, $N_z$, and $N_t$ denote the numbers of lateral pixels, axial pixels, and temporal samples, respectively. For notational clarity:
\begin{itemize}
 \item $\boldsymbol{\mathcal{X}}_{:,\,:,\,k} \in \mathbb{R}^{N_x \times N_z}$ denotes the spatial distribution of the source activity at time instant $t_k$.
 \item $\boldsymbol{\mathcal{X}}_{i,\, j,\, :} \in \mathbb{R}^{N_t}$ denotes the temporal waveform at spatial position $(x_i, z_j)$, for $i = 1, \ldots, N_x$ and $j = 1, \ldots, N_z$.
 \item $\boldsymbol{\mathcal{X}}_{i,\, j,\, k} \in \mathbb{R}$ represents the instantaneous pressure emitted by the bubbles at position ($x_i, z_j$), and time instant $t_k$.
\end{itemize}
Using a flattened index $n \in \{1, \ldots, N\}, N = N_x N_z$ linked to coordinate $(i_n, j_n)$, the datacube can be vectorized as:
\begin{equation}
 \mathbf{x} \in \mathbb{R}^{N N_t} =
 \begin{bmatrix}
 \mathbf{x}_1^\top & \cdots & \mathbf{x}_n^\top & \cdots & \mathbf{x}_N^\top
 \end{bmatrix}^\top,
\end{equation}
where $\mathbf{x}_n \in \mathbb{R}^{N_t}$ is the vector form of the temporal waveform located at position $\vec{\mathbf{r}}_n = (x_n, 0, z_n)$. 

\subsection{Linear Forward Model for Passive Acoustic Mapping}
\label{subsec:forward}

Model-based methods rely on a forward model describing the physical relationship between the observed data and the underlying source distribution. Hence, we adopt this perspective to originally formulate beamforming for PAM as a linear model-based inverse problem, under the assumption of the following linear forward model:
\begin{equation}
 \mathbf{y} = \mathbf{A} \mathbf{x} + \boldsymbol{\eta},
 \label{eq:forward_model}
\end{equation}
where $\mathbf{A} \in \mathbb{R}^{N_m N_t \times N N_t}$ denotes a linear operator that relates the recorded RF signal $\mathbf{y} \in \mathbb{R}^{N_m N_t}$, to the spatiotemporal distribution of cavitation activity $\mathbf{x} \in \mathbb{R}^{N N_t}$, and $\boldsymbol{\eta} \in \mathbb{R}^{N_m N_t}$ accounts for additive acquisition Gaussian noise.

Assuming a homogeneous, non-attenuating medium with negligible sensor directivity and system response, the acoustic propagation of a wave recorded by the $m^\text{th}$ sensor and emitted from the $n^\text{th}$ pixel is fully determined by the acoustic time-of-flight, i.e., the time required for the wavefront to reach the sensor at location $\vec{\mathbf{r}}_m$ after propagating from pixel at position $\vec{\mathbf{r}}_n$.
Hence, we define the operator $\mathbf{A}$ exclusively in terms of the acoustic time-of-flight determined by the geometric configuration of the acquisition setup. 
Under a far-field wavefront propagation model, which implies planar wavefront behavior, the discrete propagation sample delay $\delta_{m,n}$ is given by:
\begin{equation}
 \delta_{m,n} := \left\lfloor \frac{\lVert \vec{\mathbf{r}}_m - \vec{\mathbf{r}}_n \rVert_2}{c} \cdot f_s \right\rceil,
 \label{eq:delay_definition}
\end{equation}
where $c$ denotes the speed of sound in the medium, assumed to be constant and known, and $\lfloor \cdot \rceil$ represents the rounding operator to the nearest integer, ensuring that the propagating wavefront aligns with a discrete temporal sample.

We model the operator $\mathbf{A}$ as a block matrix composed of $N_m \times N$ blocks $\mathbf{A}_{m,n} \in \mathbb{R}^{N_t \times N_t}$, relating the $m^\text{th}$ sensor to the $n^\text{th}$ spatial position over $N_t$ discrete temporal samples as:
\begin{equation}
\mathbf{A} := 
\left[
 \begin{array}{ccccc}
 \mathbf{A}_{1,1} & \ldots & \mathbf{A}_{1,n} & \ldots & \mathbf{A}_{1,N} \\
 \vdots & \ddots & \vdots & \ddots & \vdots \\
 \mathbf{A}_{m,1} & \ldots & \mathbf{A}_{m,n} & \ldots & \mathbf{A}_{m,N}\\
 \vdots & \ddots & \vdots & \ddots & \vdots \\
 \mathbf{A}_{N_m,1} & \ldots & \mathbf{A}_{N_m,n} & \ldots & \mathbf{A}_{N_m,N}
 \end{array}
\right].
\label{eq:forwardmatrix}
\end{equation}
The signal recorded at the $m^\text{th}$ sensor, $\mathbf{y}_m$, is then expressed as the superposition of contributions from sources at all pixels:
\begin{equation}
 \mathbf{y}_m = \sum_{n=1}^N \mathbf{A}_{m,n} \mathbf{x}_n.
\end{equation}
In passive acquisition, a source located at the position $n$ cannot theoretically contribute to the signal recorded by sensor $m$ before the corresponding propagation delay $\delta_{m,n}$. Once this delay has elapsed, the source at position $n$ remains a persistent potential contributor to the signal recorded by sensor $m$ during subsequent temporal samples. 
This property is exploited to construct each block $\mathbf{A}_{m,n}$ as an identity matrix whose main diagonal is shifted downward by $\delta_{m,n}-1$ rows. 
In this form, the first nonzero entry appears in row $\delta_{m,n}$, column $1$, and subsequent ones follow along the subdiagonal. Equivalently:
\begin{equation}
\mathbf{A}_{m,n}[k_1,k_2] =
\begin{cases}
1, & k_1 = k_2 + \delta_{m,n} - 1, \\[4pt]
0, & \text{otherwise},
\end{cases}
\label{eq:blocks}
\end{equation}
for $k_1,k_2 = 1,\dots,N_t.$ Thus, each block $\mathbf{A}_{m,n} \in \{ 0,1\}^{N_t \times N_t}$
corresponds to a sparse binary matrix.

\input{figures/figure2}
\input{tables/table1}

Figure~\ref{fig:RF_Comparison} shows a toy example of constructing the operator $\mathbf{A}$ from the delay samples $\delta_{m,n}$, which link each sensor $m$ to each pixel $n$ and are stored in $\boldsymbol{\Delta} \in \mathbb{R}^{N_m \times N}$. A zoomed view illustrates the downward diagonal shifts imposed by the sample delays, highlighting its structured and sparse nature.

\subsection{Regularized Inversion for Passive Acoustic Mapping}
\label{subsec:inverse}
The proposed linear forward model introduced in Sec.~\ref{subsec:forward}, serves as the foundation for TD-LM-PAM framework via regularized inversion strategies. The reconstruction of the spatiotemporal cavitation activity signal, $\mathbf{x}$, is then formulated as the following inverse problem:
\begin{equation}
\underset{\mathbf{x} \in \mathbb{R}^{N N_t}}{\text{minimize}} 
\quad 
\frac{1}{2} \| \mathbf{y} - \mathbf{A} \mathbf{x} \|_2^2 
+ \lambda \, \mathcal{R}(\mathbf{x}),
\end{equation}
where the first term represents the data fidelity term, ensuring agreement with the recorded RF signals, and the second term is the regularization function, encoding prior knowledge about the structure of the cavitation activity signal, addressing the ill-posedness of the inverse problem. The parameter $\lambda$ balance the influence of the regularization term.

In this work, we explore multiple regularization strategies, selected to reflect physical and statistical characteristics of cavitation activity. To name,
\subsubsection{Sparsity (Sp) Prior} It promotes the recovery of sparse signals under the assumption that only a few spatial locations contain cavitation activity contributing significantly to the recorded RF signal~\cite{crake2018passive}. Moreover, the time domain formulation captures the transient nature of cavitation, where sources may appear and disappear depending on the considered cavitation regime. This leads to sparsity being enforced not only across spatial dimensions but also over time.
\subsubsection{Total-Variation (TV) Prior} It encourages the recovery of smooth coherent regions, considering that cavitation microbubbles are spatially localized and form compact clouds with smooth interiors and sharp boundaries~\cite{lachambre2024inverse}.
\subsubsection{Regularization by Denoising (ReD)} It incorporates denoisers as implicit priors to guide the reconstruction process. Such ReD approach is useful to promote realistic, high-quality reconstructions, enhancing source localization while effectively suppressing noise and artifacts~\cite{by2021denoiser}.

Specifically, we developed three approaches that leverage individual and combined regularizations, exploiting the complementary strengths of sparsity, smoothness, and denoising priors. The mathematical formulation for each approach is summarized in Table~\ref{tab:regularizers}.

These inverse problems can be solved using well-established optimization techniques with appropriate proximal operators tailored to each regularization. Specifically, the approach promoting sparsity is solved using the Fast Iterative Shrinkage-Thresholding Algorithm (FISTA)~\cite{beck2009fast}, which efficiently handles the $\ell_1$-norm term. For the combined regularization problems, we employ the Alternating Direction Method of Multipliers (ADMM)~\cite{boyd2011distributed}, which enables flexible splitting of the optimization terms. In particular, the Block-Matching and 4D filtering (BM4D) algorithm~\cite{maggioni2012video} is used as the implicit denoiser in the sparsity + ReD approach. The Supplementary Material for this paper provides the pseudo-code of the implemented algorithms, whose source code is publicly available at: \href{https://github.com/TatianaGelvez/TD-LM-PAM}{https://github.com/TatianaGelvez/TD-LM-PAM}.

After estimating the signal $\hat{\mathbf{x}}$, we construct a beamformed 2D spatial map $\mathbf{X} \in \mathbb{R}^{N_x \times N_z}$ representing the power of the cavitation activity, by summing the squared values for each spatial position across the temporal dimension, i.e., 
\begin{equation}
\mathbf{X}_{i_n,j_n} = \sum \hat{\mathbf{x}}_n^2,
\end{equation}
where the index tuple $(i_n, j_n) \leftrightarrow n$ corresponds to the 2D coordinates associated with the flattened index $n$.

\section{Experiments and Results}
\label{sec:Experiments}
This section presents the experiments conducted to evaluate the proposed TD-LM-PAM framework. The performance is assessed over simulated scenarios, including point sources and microbubble clouds. 
\subsection*{Comparative Benchmarking Methods}
For benchmarking purposes, the three proposed approaches, TD-LM-PAM$_\text{Sp}$, TD-LM-PAM$_\text{SpTV}$, and TD-LM-PAM$_\text{SpReD}$, summarized in Table~\ref{tab:regularizers}, are evaluated against state of the art beamforming techniques operating in both the time and frequency domains, including TD-DAS~\cite{gyongy2009passive}, FD-DAS~\cite{haworth2016quantitative}, FD-RCB~\cite{lu2018passive}, FD-CMF-ElNet~\cite{lachambre2024inverse}, and FD-CMF-SpTV~\cite{lachambre2024inverse}.

In all cases, the observations are corrupted by Gaussian noise corresponding to an SNR of $10~\mathrm{dB}$. For the frequency-domain (FD) methods, the hyperparameters are chosen as described in~\cite{lachambre2024inverse}, using $K = 130$ and an overlap of $90\%$ for estimating the CSM. For the proposed TD-LM-PAM approaches, the hyperparameters $\lambda$, $\gamma$, and $\mu$ are selected as detailed in the Supplementary Material of this paper.

\subsection*{Quantitative Evaluation Metrics}
The performance across the simulated scenarios was quantitatively assessed using well-established metrics detailed below.

For point source configurations with single microbubbles, the axial and lateral lobes were characterized using the Full Width at Half Maximum (FWHM) metric, expressed in millimeters [mm]~\cite{alomari2017plane}. The localization accuracy was assessed as the mean Euclidean distance between the true and estimated positions of each cavitation microbubble, expressed in millimeters [mm]. Specifically, the estimated positions were defined as the points of maximum amplitude in the estimated envelope. Finally, the separation capability between two sources was quantified using the peak-to-center intensity difference (PCID) metric~\cite{laroche2020inverse}. This metric measures the intensity difference, expressed in decibels [dB], between the minimum value along the line connecting two sources and their maxima, and is defined as:

\begin{equation}
\mathrm{PCID} = 20 \log_{10} \left( \frac{I_{\text{min}}}{I_{\text{max}}} \right),
\end{equation}
where $I_{\text{max}}$ represents the amplitude at the maxima associated with the two sources, and $I_{\text{min}}$ denotes the minimum amplitude between them. Note that, in sparse images, where the minimum between peaks approaches zero, the PCID can theoretically reach minus infinity. In this work, two point sources are considered unresolved when PCID $>-6$~dB~\cite{xu2025comparative}.
 
For cloud configurations containing grouped microbubbles, the classical Contrast-to-Noise Ratio (CNR)~\cite{liebgott2016plane, rodriguez2019generalized} was employed to quantify the separability between the cavitation cloud and the surrounding background. It is calculated as:

\begin{equation}
\text{CNR} = 20 \log_{10} \left( \frac{|\mu_i - \mu_o|}{\sqrt{\sigma_i^2 + \sigma_o^2}} \right),
\end{equation}
where $\mu_i$, $\mu_o$ and $\sigma_i$, $\sigma_o$ are the means and standard deviations of the signal and noise zones, respectively, taken from the estimated power map of the cavitation activity. The signal zone corresponds to the area inside the cavitation cloud, while the noise zone is defined as a 2~mm margin surrounding the cloud.

To evaluate the fidelity of estimated cavitation cloud shapes, we used the Dice coefficient computed at $-3\,\mathrm{dB}$. True positives (TP) are defined as pixels with power greater than or equal to $-3\,\mathrm{dB}$ located within the signal zone. False negatives (FN) are pixels within the signal zone with power below $-3\,\mathrm{dB}$. False positives (FP) are pixels with power greater than or equal to $-3\,\mathrm{dB}$ located in the noise zone. The Dice coefficient, which quantifies the overlap between the detected and ground-truth cavitation zones, is defined as:

\begin{equation}
\mathrm{Dice} = \frac{2 |X \cap Y|}{|X| + |Y|}, 
\end{equation}
where $X$ represents the binary image corresponding to the signal power zone (i.e., TP + FN), and $Y$ denotes the binary image corresponding to all pixels above $-3\,\mathrm{dB}$ (i.e., TP + FP).

\subsection{Validation of the Linear Forward Model}
To validate the consistency of the proposed forward model and its ability to capture time-delay effects during RF signal acquisition in a linear array, we compare its output with reference observations obtained using the simulation scheme presented in~\cite{lauterborn2010physics}. 

The comparison is conducted in a noise-free scenario using a discretized setup to minimize numerical errors. The resulting error, expressed as the normalized mean square error (NMSE) between the observations obtained from the reference simulation scheme, $\mathbf{y}_{\text{sim}}$, and those obtained from the proposed operator, $\mathbf{y}_{\mathbf{A}}$, is evaluated over 100 experiments using different input images. Across these realizations, the NMSE exhibits a mean of $1.83 \times 10^{-15}$ and a standard deviation of $3.11 \times 10^{-16}$, confirming that the operator is consistent with the literature simulation scheme.

\subsection{Quantitative Results on Point Source Configuration}
This experiment follows the two point-source configurations described in~\cite{lachambre2024inverse} to assess the performance of the proposed TD-LM-PAM framework in separating closely spaced sources. Specifically, we evaluate the spatial resolution and source separation capabilities, including axially distributed point sources, which typically represent the most challenging scenario.

The first configuration consists of two laterally distributed inertial cavitation point-sources located at $(-5, 72)$~mm and $(-3, 72)$~mm in the $(x, 0, z)$ plane. The second configuration involves two axially distributed inertial cavitation point sources positioned at $(-3, 64)$~mm and $(-3, 72)$~mm. The RF signals are recorded over a duration of $N_t = 200~\mu\text{s}$ using a linear probe equipped with $N_m = 128$ sensors. Note that, the TD methods use only 20\% (40 $\mu$s) of the observed RF signal, whereas the FD methods use the entire signal (200 $\mu$s), required to accurately estimate the CSM.

\input{tables/table3}
\input{tables/table4}
\input{figures/figure3Arxiv}

Tables~\ref{table:PointSourceLateral} and~\ref{table:PointSourceAxial} present the average performance computed over fifty independent realizations of the same scenario for the lateral and axial configurations, respectively. The best result for each metric is highlighted in bold, while the second-best result is underlined. 

The results demonstrate that TD-LM-PAM, with appropriate regularization, outperforms state-of-the-art methods. In particular, the sparsity-regularized approach (TD-LM-PAM$_{\text{Sp}}$) proves to be the most effective. This behavior is expected, as sparsity serves as a prior consistent with the assumption of point-like sources emitting signals, whereas the smoothing and denoising priors employed by the SpTV and SpRed approaches are less suitable for isolated, point-like sources. TD-LM-PAM$_{\text{Sp}}$ achieves the narrowest axial lobes, as evidenced by the smallest FWHM (below or equal to 1~mm), highlighting its superiority for axially distributed point sources. Moreover, despite using significantly fewer measurements, TD-LM-PAM$_{\text{Sp}}$ achieves comparable accuracy in localizing microbubble positions, as reflected by the position error, typically exhibiting a similar average performance but with slightly higher variance compared to the best result.

\input{figures/figure4Arxiv}
\input{figures/figure5Arxiv}
\input{figures/figure6Arxiv}

For a qualitative comparison, Figs.~\ref{fig:BubbleMapsLateral} and~\ref{fig:BubbleMapsAxial} depict representative examples of the reconstructed spatial maps, illustrating the reconstructed microbubble distributions for each method and configuration. Furthermore, Figs.~\ref{fig:ProfilesLateral} and~\ref{fig:ProfilesAxial} show the corresponding lateral and axial profiles, allowing for a detailed evaluation of both resolution and localization accuracy. Notably, the axial profiles demonstrate that the lobe generated by TD-LM-PAM$_{\text{Sp}}$ is the narrowest, underscoring its superior capability for source discrimination compared to TD and other FD state-of-the-art approaches. 
In addition, we observed that the main localization errors for axially distributed point sources predominantly arise from sources located at greater depths, where the problem is more ill-posed and the accuracy of their reconstruction is consequently limited.

\subsection{Quantitative Results on Cloud Configuration}
\label{subsec:cloudsExperiment}
Building on the point-source scenario, we next evaluate the TD-LM-PAM framework under a more challenging cloud configuration, in which individual microbubbles are grouped to better approximate realistic conditions. This experiment follows the same setup as the circular cloud scenario described in~\cite{lachambre2024inverse}, consisting of a circular source with a diameter of 2~mm, centered at $(-7,\,70)$~mm, with a density of 100~point~sources/mm$^2$.

\input{figures/figure7Arxiv}

Table~\ref{tab:CloudResults} reports the quantitative results showing that TD-LM-PAM consistently outperforms benchmark methods, yielding sharper reconstructions. In particular, based on CNR, TD-LM-PAM$_{\text{SpRed}}$ proves to be the most effective method, enabling the recovery of dense microbubble clouds while preserving their shape and effectively suppressing surrounding noise. Overall, TD-LM-PAM with appropriate regularizers demonstrates the ability to preserve fine-scale structural information even in dense regions.

Figure~\ref{fig:CloudMaps} shows typical examples of the reconstructed spatial maps, visually evidencing that the proposed approach provides satisfactory estimation while avoiding artifacts and preserving contrast. Notably, TD-LM-PAM$_\text{SpRed}$ preserves the cloud shape, as further supported by the DICE metric. In contrast, the methods TD-LM-PAM$_\text{sp}$ and TD-LM-PAM$_\text{SpTV}$ provide comparable results to the ones obtained with FD methods.

\input{tables/table5}

\section{Conclusions}
We proposed TD-LM-PAM, a time-domain linear model framework for passive acoustic mapping that formulates beamforming as a regularized inverse problem. Unlike existing methods, our approach introduces for the first time a linear forward model consistent with the acquisition dynamics, enabling the use of diverse regularized inverse approaches. Experiments with point sources and microbubble clouds demonstrate improved axial resolution, higher contrast and comparable localization accuracy compared to state-of-the-art TD and FD methods. In perspective, the performance is competitive to FD methods, with the advantage of using only 20\% of the data. Besides, FD approaches typically impose a fixed reconstruction frequency. These results highlight the potential of TD-LM-PAM as an efficient and versatile tool for cavitation activity beamforming.

\section*{Acknowledgment}

\bibliographystyle{IEEEtran} % o cualquier estilo que prefieras, ejemplo: plain, unsrt, abbrv
\bibliography{references} % nombre del archivo .bib sin la extensión

\end{document}

%% file: figures/figure1.tex
\begin{figure}[t!]
    \centering
    \begin{adjustbox}{width=1\linewidth}
        \begin{tikzpicture}

%%%Image%%%%
\node[anchor=center, inner sep=0] (img)  at (0,0) {\includegraphics[width=1\linewidth]{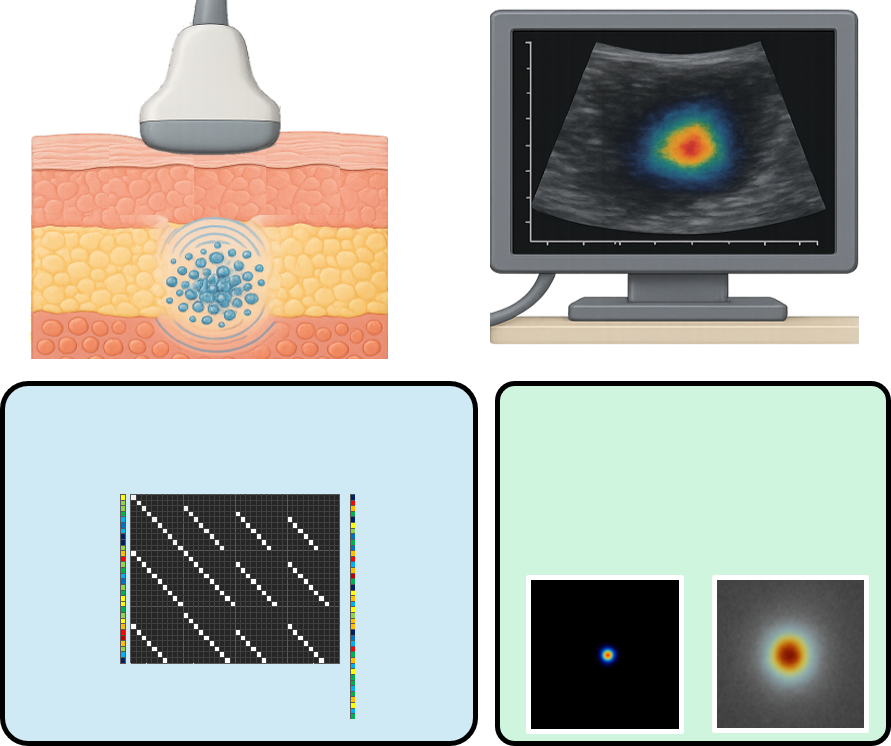}};

%%%Labels Text%%%%%
\node[anchor = center, inner sep=1pt, fill=white] at (0.0\linewidth,3.85) {\small{\textbf{(a) Passive Acoustic Mapping}}};

\node[anchor = center, inner sep=1pt, align=center] at (-0.38\linewidth,2.9) {\small{Probe}};

\draw[->] (-0.27\linewidth,2.3) -- (-0.38\linewidth,2.7);

\node[anchor = center, inner sep=1pt, align=center] at (-0.35\linewidth,1.4) {\small{Microbubbles}};

\draw[->] (-0.27\linewidth,0.85) -- (-0.33\linewidth,1.25);

\node[anchor = center, inner sep=1pt, text width=3.2cm, align=center]  at (0.25\linewidth,0.05) {\small{Cavitation maps}};

\draw[->, thick] (0.25\linewidth,1.1) -- (0.25\linewidth,0.2);

%%%Labels Math%%%%
\node[anchor = center, inner sep=0] at (-0.22\linewidth,-0.5) {\small{\textbf{(b) Linear Forward Model}}};

\node[anchor = center, inner sep=0] at (-0.22\linewidth,-0.95) {\small{$\mathbf{y} = \mathbf{A}\mathbf{x} + \boldsymbol{\eta} $}};

\node[anchor = center, inner sep=0] at (-0.355\linewidth,-3.25) {\small{$\mathbf{y}$}};

\node[anchor = center, inner sep=0] at (-0.245\linewidth,-3.25) {\small{$\mathbf{A}$}};

\node[anchor = center, inner sep=0] at (-0.08\linewidth,-3.25) {\small{$\mathbf{x}$}};

\node[anchor = center, inner sep=0] at (0.265\linewidth,-0.5) {\small{\textbf{(c) Regularized Inversion}}};

\node[anchor = center, inner sep=0] at (0.27\linewidth,-1.25) {\footnotesize{$\min_x \Vert \mathbf{y} - \mathbf{A}\mathbf{x}\Vert_2^2 +  \mathcal{R}(\mathbf{x}) $}};

\node[anchor = center, inner sep=0] at (0.18\linewidth,-1.8) {\small{Sparsity}};

\node[anchor = center, inner sep=0] at (0.38\linewidth,-1.8) {\small{Smoothness}};

\end{tikzpicture}
\end{adjustbox}
\caption{Time-Domain Linear Model for Passive Acoustic Mapping (TD-LM-PAM) framework.
(a) Cavitation monitoring, where a probe passively records acoustic emissions from clouds of microbubbles, followed by beamforming to display cavitation maps.
(b) Forward model relying in a linear operator $\mathbf{A}$, linking the spatiotemporal distribution  of cavitation activity, $\mathbf{x}$, to the recorded radio-frequency signals, $\mathbf{y}$.
(c) General scheme of regularized inversion enabling the incorporation of prior knowledge, such as sparsity or smoothness.}
\label{fig:PAM}
\end{figure}

%% file: figures/figure2a.tex
\begin{figure}[b!]
    \centering
    \begin{adjustbox}{width=1\linewidth}
        \begin{tikzpicture}
            % Imagen
            \node[anchor=north west, inner sep=0] (img) 
                at (0,0) {\includegraphics[width=1\linewidth]{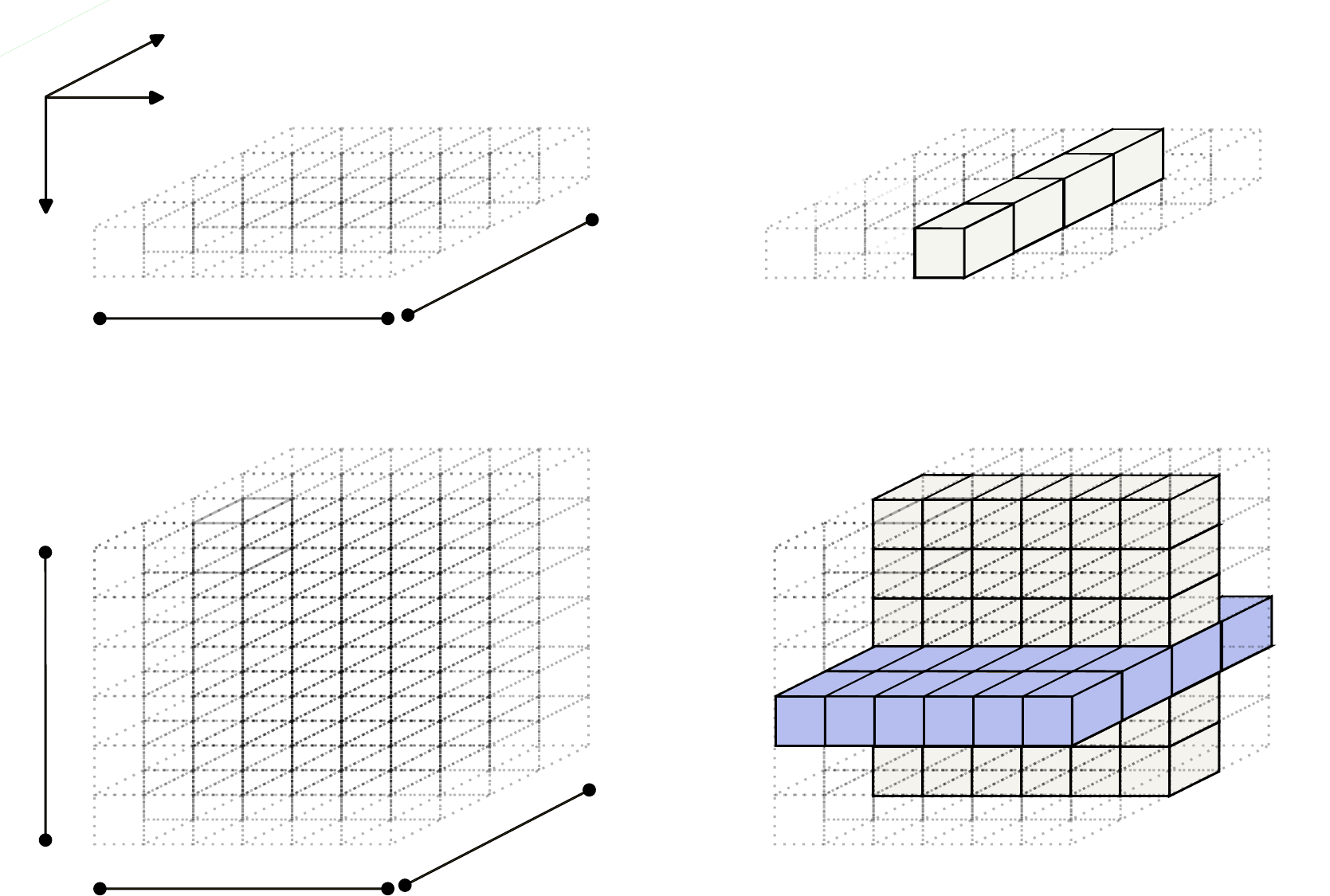}};
                
\draw[ anchor = north west, <-, ultra thick, color={rgb,255:red,0; green,194; blue,160}]
(0.72\linewidth, -0.8) to (0.77\linewidth, -1.1);

\draw[ anchor = north west, <-, ultra thick, color={rgb,255:red,0; green,194; blue,160}]
(0.72\linewidth, -2.8) to (0.77\linewidth, -3.1);

\draw[ anchor = north west, ->, ultra thick, color={rgb,255:red,0; green,194; blue,160}]
(0.62\linewidth, -5.1) to (0.57\linewidth, -5.4);

\node [anchor = north west] at (0.57\linewidth,-0.3) {\small{$\mathbf{y}_m \in \mathbb{R}^{N_t}, m = 1, \ldots N_m$}};

\node [anchor = north west] at (0.47\linewidth,-2.3) {\small{$\boldsymbol{\mathcal{X}}_{:,:,k} \in \mathbb{R}^{N_x \times N_z}, k = 1, \ldots N_t$}};

\node [anchor = north west] at (0.47\linewidth,-5.4) {\small{$\boldsymbol{\mathcal{X}}_{:,j,:} \in \mathbb{R}^{N_x \times N_t}, j = 1, \ldots N_z$}};

\node[anchor = north west] at (0.15\linewidth, 0.25) {\small{$(x,z,t):$} \small{lateral, axial, and temporal dimensions.}};
\node [anchor = north west] at (0.125\linewidth,-0.4) {\small{$x$}};
\node [anchor = north west] at (0.035\linewidth,-1.2) {\small{$z$}};
\node [anchor = north west] at (0.125\linewidth,0.2) {\small{$t$}};
\node [anchor = north west] at (0.20\linewidth,-0.3) {\small{$\mathbf{Y}:$ RF signal}};
\node [anchor = north west, draw = black, fill = white, inner sep=0.06em] at (0.16\linewidth,-2.0) {\footnotesize{$N_m$}};
\node [anchor = north west, draw = black, fill = white, inner sep=0.06em] at (0.35\linewidth,-1.65) {\footnotesize{$N_t$}};

\node [anchor = north west] at (0.10\linewidth,-2.5) {\small{$\boldsymbol{\mathcal{X}}$} cavitation datacube};
\node [anchor = north west, draw = black, fill = white, inner sep=0.06em] at (0.015\linewidth,-4.5) {\footnotesize{$N_z$}};
\node [anchor = north west, draw = black, fill = white, inner sep=0.06em] at (0.16\linewidth,-5.8) {\footnotesize{$N_x$}};
\node [anchor = north west, draw = black, fill = white, inner sep=0.06em] at (0.35\linewidth,-5.4) {\footnotesize{$N_t$}};

\end{tikzpicture}
\end{adjustbox}
 \caption{Mathematical notation scheme.  $\mathbf{Y} \in \mathbb{R}^{N_m \times N_t}$ denotes the RF signals, and $\boldsymbol{\mathcal{X}} \in \mathbb{R}^{N_x \times N_z \times N_t }$ represents the cavitation spatiotemporal datacube.}
\label{fig:notation}
\end{figure}
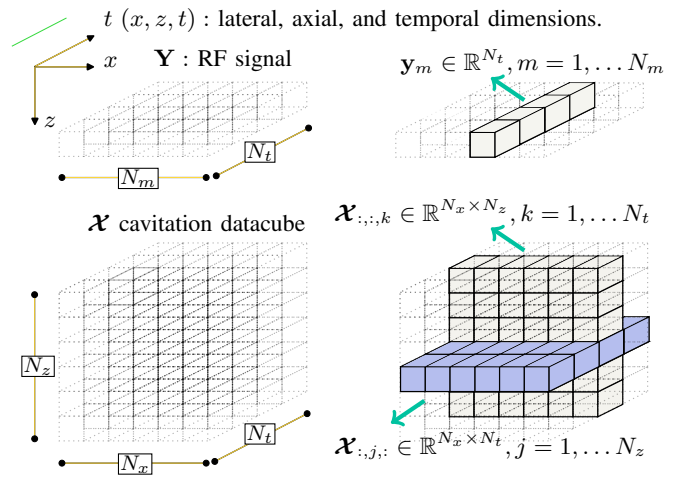

%% file: figures/figure2.tex
\begin{figure}[b!]
    \centering
    \begin{tikzpicture}
        \node[anchor = north west, inner sep=0pt] (img) {\includegraphics[width=1\linewidth]{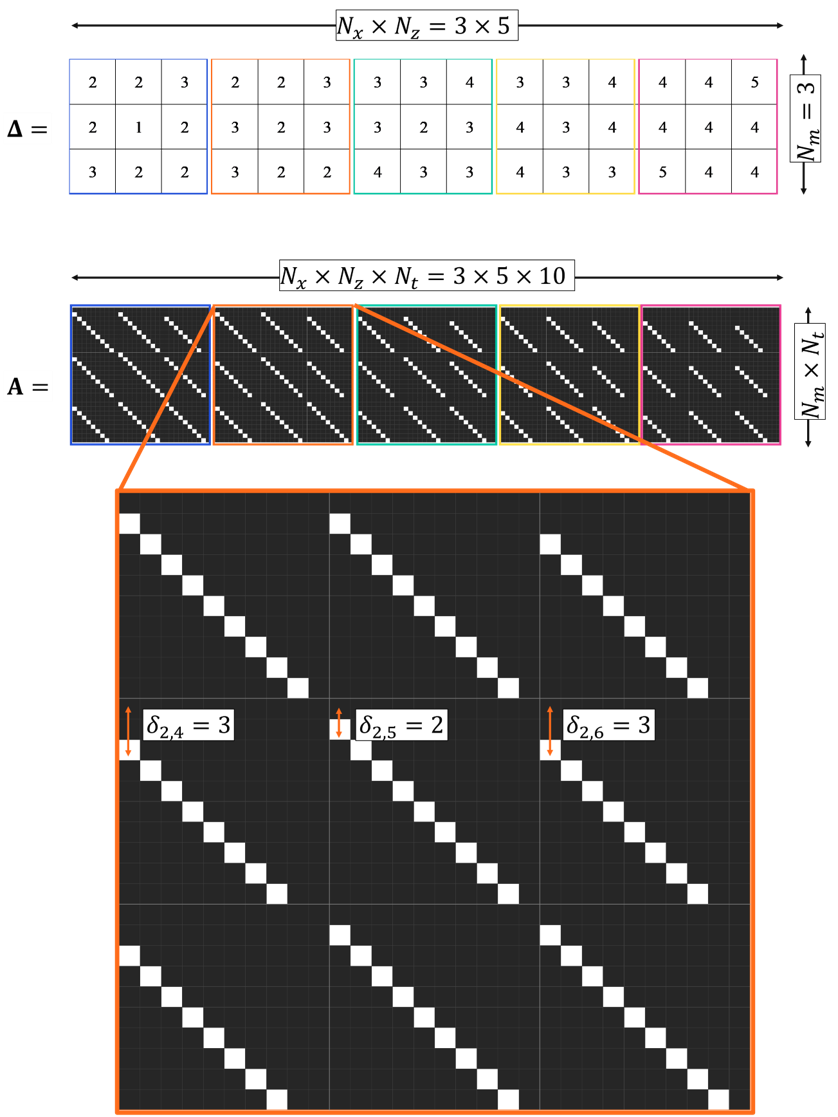}};
        % Nodo de texto en la esquina inferior izquierda

\node[anchor=north west, font=\scriptsize, fill = white, draw=black, inner sep = 0.25em]     at (0.18\textwidth, -0.08) {$N_x \times N_z = 3 \times 5$};

%\node[anchor=north west, font=\scriptsize, fill = white, inner sep = 0.03em]  at (0.31\textwidth, 0.05) {\textbf{(b) Kernel Patterns} $\Omega_{:,k}$};

 %\node[anchor=north west, font=\scriptsize]  at (0.085\textwidth, -4.3) {$1$};

%\node[anchor=north west, font=\scriptsize] at (0.18\textwidth, -4.3) {$0$};

%\node[anchor=north west, font=\scriptsize] at (0.283\textwidth, -4.3) {$k = 93$};
   
%\node[anchor=north west, font=\scriptsize]  at (0.355\textwidth, -4.3) {$k = 137$};

%\node[anchor=north west, font=\scriptsize]  at (0.43\textwidth, -4.3) {$k=207$};

    \end{tikzpicture}
    \caption{Linear forward operator toy example. The image plane consists of $N_x \times N_z = 3 \times 5$ pixels, observed over $N_t = 10$ time instants and recorded with a probe of $N_m = 3$ sensors. The structure of the operator $\mathbf{A}$ is determined by the sample delays stored in the matrix $\boldsymbol{\Delta}$. The zoomed-in figure shows the sub-blocks corresponding to the relationship between the third sensor and the lateral pixels for the second axial pixel.}
    \label{fig:RF_Comparison}
\end{figure}

%% file: tables/table1.tex
\begin{table*}[b!]
\caption{Regularized Optimization Approaches Formulation}
\centering
\renewcommand{\arraystretch}{1.9} % Espaciado vertical
\begin{tabularx}{\textwidth}{>{\centering\arraybackslash}m{0.09\textwidth} >{\raggedright\arraybackslash}m{0.13\textwidth} >{\raggedright\arraybackslash}X}
\hline
\multicolumn{1}{c}{\textbf{Approach}} & \multicolumn{1}{c}{\textbf{Abbreviation}} & \multicolumn{1}{c}{\textbf{Inverse Problem Formulation}} \\
\hline \hline

{\centering \textbf{Sparsity}} & TD-LM-PAM$_{\text{Sp}}$ &
$\hat{\mathbf{x}} \in \underset{\mathbf{x} \in \mathbb{R}^{N  N_t}}{\arg\min}
\left\{ 
\frac{1}{2} \left\| \mathbf{y} - \mathbf{A} \mathbf{x} \right\|_2^2 
+ \lambda \left\| \mathbf{x} \right\|_1
\right\}$,  $\ell_1$-norm: $\|\mathbf{x}\|_1 = \sum_{i,j,k} |\mathcal{X}_{i,j,k}|$. \\
\hline

\parbox[c][\height][c]{\linewidth}{\centering \vspace{6mm} \textbf{Sparsity + TV}} & \vspace{6mm}  TD-LM-PAM$_{\text{SpTV}}$&
$\hat{\mathbf{x}} \in \underset{\mathbf{x} \in \mathbb{R}^{N N_t}}{\arg\min}
\left\{ 
\frac{1}{2} \left\| \mathbf{y} - \mathbf{A} \mathbf{x} \right\|_2^2 
+ \lambda \left\| \mathbf{x} \right\|_1 
+ \gamma \left\| \mathbf{x} \right\|_{\mathrm{TV}}
\right\}$, \par TV-norm: 
$\|\mathbf{x}\|_{\mathrm{TV}} := 
\sum_{i,j,k} \Big( |\mathcal{X}_{i+1,j,k} - \mathcal{X}_{i,j,k}| 
+ |\mathcal{X}_{i,j+1,k} - \mathcal{X}_{i,j,k}| 
+ |\mathcal{X}_{i,j,k+1} - \mathcal{X}_{i,j,k}| \Big) = \left\| \mathbf{D}\mathbf{x} \right\|_1$. Here, $\mathbf{D}$ denotes the discrete differences operator. \\
\hline

\parbox[c][\height][c]{\linewidth}{\centering \textbf{Sparsity + ReD}} & TD-LM-PAM$_{\text{SpReD}}$ &
$\hat{\mathbf{x}} \in \underset{\mathbf{x} \in \mathbb{R}^{N  N_t}}{\arg\min}
\left\{ 
\frac{1}{2} \left\| \mathbf{y} - \mathbf{A} \mathbf{x} \right\|_2^2 
+ \lambda \left\| \mathbf{x} \right\|_1 
+ \mu \mathcal{R}_{\mathrm{D}}(\mathbf{x})
\right\}$,  ReD-term: 
$\mathcal{R}_{\mathrm{D}}(\mathbf{x}) = \frac{1}{2} \mathbf{x}^\top \left( \mathbf{x} - f(\mathbf{x}) \right)$. \\
\hline
\end{tabularx}
\label{tab:regularizers}
\end{table*}

%% file: tables/table3.tex
\begin{table}[b!]
\caption{Average Quantitative Performance over 50 Replicas of Two Laterally Distributed Point Sources}
\centering
\footnotesize
\renewcommand{\arraystretch}{1.25} 
\begin{tabularx}{1\linewidth}{X|c|c|c|c}
\hline
\makecell[c]{\textbf{Method}} & 
\makecell[c]{\textbf{Axial}\\\textbf{FWHM}\\\textbf{[mm]}} & 
\makecell[c]{\textbf{Lateral}\\\textbf{FWHM}\\\textbf{[mm]}} & 
\makecell[c]{\textbf{Position}\\\textbf{Error}\\\textbf{[mm]}} & 
\makecell[c]{\textbf{Separation}\\\textbf{Power}\\\textbf{[dB]}} \\ \hline \hline
 \textbf{FD-DAS} \makecell[c]{\rule{0pt}{5ex}}      &  $7.4 \,(0.5)$   &  $0.61 \,(0.01)$  &  $1.1 \,(0.4)$   &  $-13.9 \,(0.7)$ \\ \hline
  \textbf{FD-RCB} \makecell[c]{\rule{0pt}{5ex}}     &  $3.3 \,(0.2)$    &  $\underline{<0.2}$            &  $\mathbf{0.10 \,(0.01)}$   &  $<-20$  \\ \hline
 \textbf{FD-CMF-ElNet}  &   $\underline{1.3 \,(0.2)}$     &  $\underline{<0.2}$            &  $\underline{0.1 \,(0.1)}$   &   $<-20$ \\ \hline
\textbf{FD-CMF-SpTV}   &  $1.4 \,(0.1)$    &  $0.44 \,(0.01)$   &  $0.2 \,(0.1)$   &  $<-20$ \\ \hline
 \textbf{TD-DAS}  \makecell[c]{\rule{0pt}{5ex}}   &  $12.6\,(2.6)$           &  $0.40\,(0.01)$             &  $0.74\,(0.04)$           &  $-13.5 (0.5)$ \\ \hline
  \textbf{TD-LM-PAM$_\text{Sp}$}     &  $\mathbf{0.8\,(0.2)}$ & $\mathbf{<0.1}$ & $0.1\,(0.2)$ &   $<-20$ \\ \hline
\textbf{TD-LM-PAM$_\text{SpTV}$}   &  $2.4\,(0.3)$ & $\underline{<0.2}$ & $0.4\,(0.1)$ &   $<-20$ \\ \hline
\textbf{TD-LM-PAM$_\text{SpRed}$}  &  $1.4\,(0.2)$ & $\underline{<0.2}$ & $0.4\,(0.2)$ &   $<-20$ \\ \hline
\end{tabularx}
\label{table:PointSourceLateral}
\end{table}

% \begin{table*}[tbp]
% \caption{Average Quantitative Performance over 50 Replicas of \\ Two Axially Distributed Point Sources}
% \centering
% \normalsize
% \renewcommand{\arraystretch}{1.25} 
% \resizebox{0.6\textwidth}{!}{
% \begin{tabularx}{0.6\linewidth}{X|c|c|c|c}
% \hline
% \makecell[c]{\textbf{Method}} & 
% \makecell[c]{\textbf{Axial}\\\textbf{FWHM}\\\textbf{[mm]} $\downarrow$} & 
% \makecell[c]{\textbf{Lateral}\\\textbf{FWHM}\\\textbf{[mm]} $\downarrow$} & 
% \makecell[c]{\textbf{Position}\\\textbf{Error}\\\textbf{[mm]} $\downarrow$} & 
% \makecell[c]{\textbf{Separation}\\\textbf{Power}\\\textbf{[dB]} $\downarrow$} \\ \hline \hline
% \textbf{FD-DAS}     &  $14.0 \,(2.0)$   &  $0.67 \,(0.01)$  &  $1.7 \,(0.3)$   &  $-1.6 \,(0.7)$ \\ \hline
% \textbf{FD-RCB}     &  $3.5 \,(0.5)$    &  $<0.2$            &  $0.6 \,(0.6)$   &  $-5.7 \, (0.6)$ \\ \hline
% \textbf{FD-CMF-ElNet}   &  $1.3 \,(0.2)$    &  $<0.2$            &  $0.3 \,(0.2)$   &   $<-20$ \\ \hline
% \textbf{FD-CMF-SpTV}    &  $1.7 \,(0.5)$    &  $0.45 \,(0.01)$   &  $0.4 \,(0.3)$   &  $<-20$ \\ \hline
% \textbf{TD-DAS}     &  $17.6\,(2.1)$           &  $0.6\,(0.01)$             &  $3.9\,(0.4)$           &  $<-20$ \\ \hline
% \textbf{TD-LM-PAM$_\text{Sp}$}     &  $\mathbf{1.2}\,(0.1)$ & $<0.1$ & $0.1\,(0.1)$ &   $<-20$ \\ \hline
% %\textbf{TD-LM-PAM$_\text{SpTV}$}   &  $\mathbf{1.2}\,(0.1)$ & $<0.1$ & $0.1\,(0.1)$ &   $<-20$ \\ \hline
% %\textbf{TD-LM-PAM$_\text{SpRed}$}  &  $1.4\,(0.1)$ & $<0.1$ & $0.1\,(0.1)$ &   $<-20$ \\ \hline
% \end{tabularx}
% }
% \label{table:PointSource}
% \end{table*}

%% file: tables/table4.tex
\begin{table}[t!]
\caption{Average Quantitative Performance over 50 Replicas of  Two Axially Distributed Point Sources}
\centering
\footnotesize
\renewcommand{\arraystretch}{1.25} 
\begin{tabularx}{1\linewidth}{X|c|c|c|c}
\hline
\makecell[c]{\textbf{Method}} & 
\makecell[c]{\textbf{Axial}\\\textbf{FWHM}\\\textbf{[mm]}} & 
\makecell[c]{\textbf{Lateral}\\\textbf{FWHM}\\\textbf{[mm]}} & 
\makecell[c]{\textbf{Position}\\\textbf{Error}\\\textbf{[mm]}} & 
\makecell[c]{\textbf{Separation}\\\textbf{Power}\\\textbf{[dB]}} \\ \hline \hline
 \textbf{FD-DAS} \makecell[c]{\rule{0pt}{5ex}}      &  $14.0 \,(2.0)$   &  $0.67 \,(0.01)$  &  $1.7\,(0.3)$   &  $-1.6 \,(0.7)$ \\ \hline
  \textbf{FD-RCB} \makecell[c]{\rule{0pt}{5ex}}     &  $3.5 \,(0.5)$    &  $<0.2$            &  ${0.6 \,(0.6)}$   &  $<-5.7(0.6)$  \\ \hline
 \textbf{FD-CMF-ElNet}  &   $\underline{1.3 \,(0.2)}$     &  $<0.2$            &  $\mathbf{0.3 \,(0.2)}$   &   $<-20$ \\ \hline
\textbf{FD-CMF-SpTV}   &  $1.7 \,(0.5)$    &  $0.45 \,(0.01)$   &  $0.4 \,(0.3)$   &  $<-20$ \\ \hline
 \textbf{TD-DAS}  \makecell[c]{\rule{0pt}{5ex}}   &  $15.7\,(0.2)$           &  $0.4\,(0.01)$ &  $0.6\,(0.1)$     &  $<-3.4 (0.8)$ \\ \hline
  \textbf{TD-LM-PAM$_\text{Sp}$}     &  $\mathbf{1.0\,(0.2)}$ & $\mathbf{<0.02}$ & $\underline{0.3\,(0.3)}$ &   $<-20$ \\ \hline
\textbf{TD-LM-PAM$_\text{SpTV}$}   &  $2.3\,(0.3)$ & $ \underline{<0.1}$ & $0.6\,(0.1)$ &   $<-20$ \\ \hline
\textbf{TD-LM-PAM$_\text{SpRed}$}  &  $1.7\,(0.3)$ & $\underline{<0.1}$ & $0.4\,(0.1)$ &   $<-20$ \\ \hline
\end{tabularx}
\label{table:PointSourceAxial}
\vspace{-6mm}
\end{table}

%% file: figures/figure3Arxiv.tex
\begin{figure}[b!]
    \centering
    \begin{tikzpicture}
        \node[anchor = north west, inner sep=0pt] (img) {\includegraphics[width=1\linewidth]{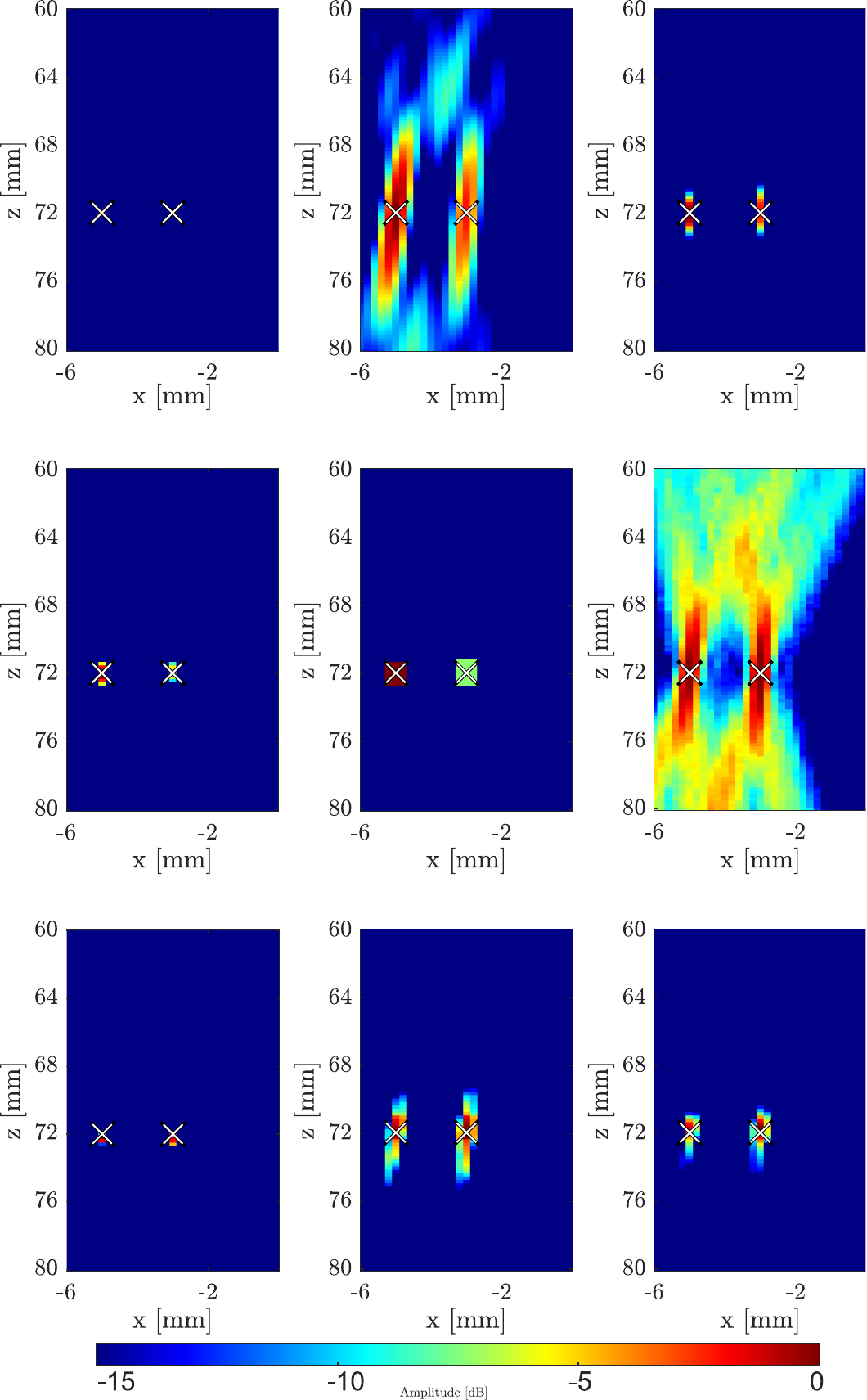}};
        \node[anchor  = north west, inner sep=0pt, align=center] 
        at (0.18\linewidth,0.2) {\footnotesize{\textbf{GT}}}; 
        \node[anchor  = north west, inner sep=0pt, align=center] 
        at (0.48\linewidth,0.2) {\footnotesize{\textbf{FD-DAS}}};
        \node[anchor  = north west, inner sep=0pt, align=center] 
        at (0.805\linewidth,0.2) {\footnotesize{\textbf{FD-RCB}}}; 
        \node[anchor  = north west, inner sep=0pt, align=center] 
        at (0.09\linewidth,-4.5) {\footnotesize{\textbf{FD-CMF-ElNet}}};
        \node[anchor  = north west, inner sep=0pt, align=center] 
        at (0.44\linewidth,-4.5) {\footnotesize{\textbf{FD-CMF-SpTV}}};
        \node[anchor  = north west, inner sep=0pt, align=center] 
        at (0.81\linewidth,-4.5) {\footnotesize{\textbf{TD-DAS}}};
        \node[anchor  = north west, inner sep=0pt, align=center] 
        at (0.09\linewidth,-9.15) {\footnotesize{\textbf{TD-LM-PAM}$_{\text{\textbf{Sp}}}$}};
         \node[anchor  = north west, inner sep=0pt, align=center] 
        at (0.44\linewidth,-9.15) {\footnotesize{\textbf{TD-LM-PAM}$_{\text{\textbf{SpTV}}}$}};
         \node[anchor  = north west, inner sep=0pt, align=center] 
        at (0.75\linewidth,-9.15) {\footnotesize{\textbf{TD-LM-PAM}$_{\text{\textbf{SpReD}}}$}};
        \node[anchor  = north west, inner sep=0.03pt, align=center, fill = white] at (0.45\linewidth,-14.1) {\footnotesize{Amplitude}}; 
        \end{tikzpicture}
        \caption{Estimated power maps of two laterally distributed inertial bubbles simulated at (-5,72)~mm and (-3, 72)~mm. %across diverse frequency and time domain methods.
        }
\label{fig:BubbleMapsLateral}
    \end{figure}
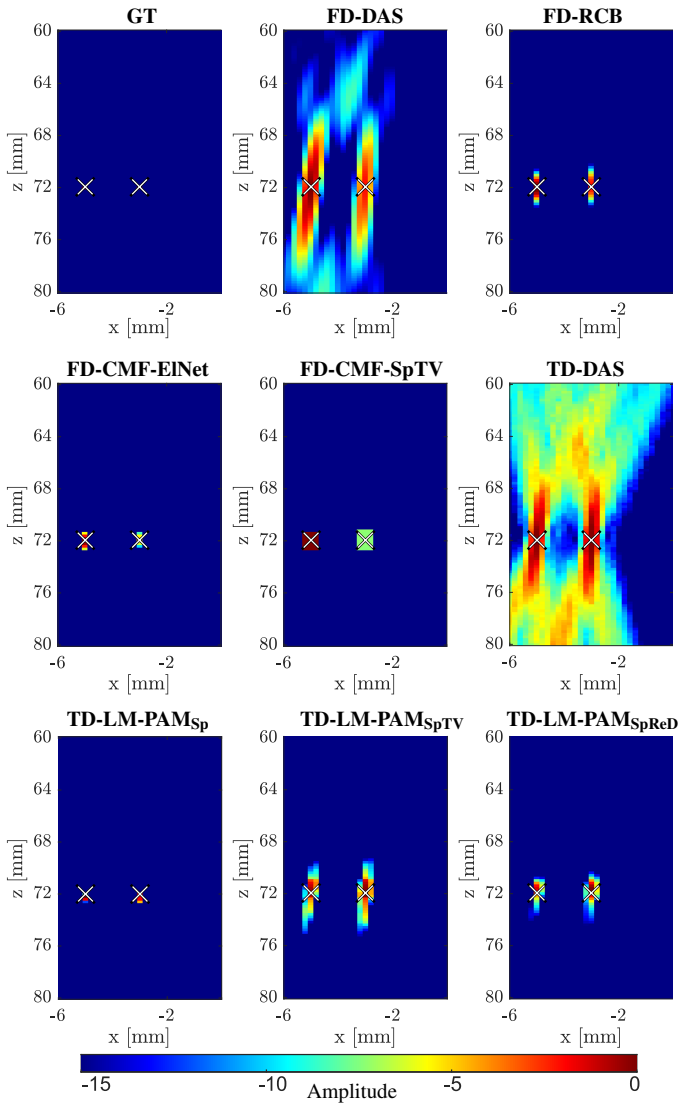

%% file: figures/figure4Arxiv.tex
\begin{figure}[b!]
    \centering
    \begin{tikzpicture}
        \node[anchor = north west, inner sep=0pt] (img) {\includegraphics[width=1\linewidth]{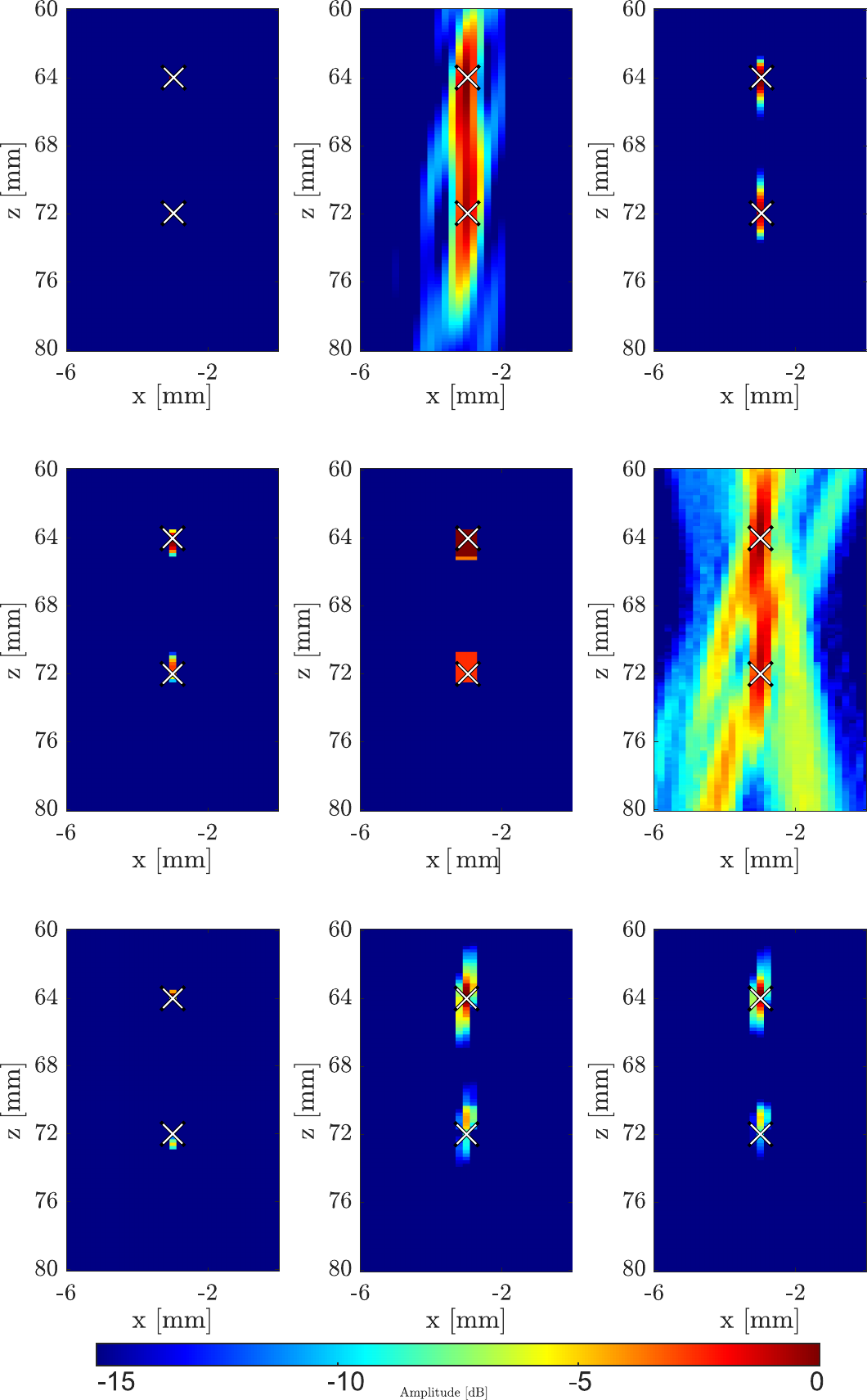}};
       \node[anchor  = north west, inner sep=0pt, align=center] 
        at (0.18\linewidth,0.2) {\footnotesize{\textbf{GT}}}; 
        \node[anchor  = north west, inner sep=0pt, align=center] 
        at (0.48\linewidth,0.2) {\footnotesize{\textbf{FD-DAS}}};
        \node[anchor  = north west, inner sep=0pt, align=center] 
        at (0.805\linewidth,0.2) {\footnotesize{\textbf{FD-RCB}}}; 
        \node[anchor  = north west, inner sep=0pt, align=center] 
        at (0.09\linewidth,-4.5) {\footnotesize{\textbf{FD-CMF-ElNet}}};
        \node[anchor  = north west, inner sep=0pt, align=center] 
        at (0.44\linewidth,-4.5) {\footnotesize{\textbf{FD-CMF-SpTV}}};
        \node[anchor  = north west, inner sep=0pt, align=center] 
        at (0.81\linewidth,-4.5) {\footnotesize{\textbf{TD-DAS}}};
        \node[anchor  = north west, inner sep=0pt, align=center] 
        at (0.09\linewidth,-9.15) {\footnotesize{\textbf{TD-LM-PAM}$_{\text{\textbf{Sp}}}$}};
         \node[anchor  = north west, inner sep=0pt, align=center] 
        at (0.44\linewidth,-9.15) {\footnotesize{\textbf{TD-LM-PAM}$_{\text{\textbf{SpTV}}}$}};
         \node[anchor  = north west, inner sep=0pt, align=center] 
        at (0.75\linewidth,-9.15) {\footnotesize{\textbf{TD-LM-PAM}$_{\text{\textbf{SpReD}}}$}};
        \node[anchor  = north west, inner sep=0.03pt, align=center, fill = white] 
        at (0.45\linewidth,-14.1) {\footnotesize{Amplitude}};  
        \end{tikzpicture}
        \caption{Estimated power maps of two axially distributed inertial bubbles simulated at (-3, 64)~mm and (-3, 72)~mm.%, across diverse frequency and time domain methods.
        }
        \label{fig:BubbleMapsAxial}
    \end{figure}
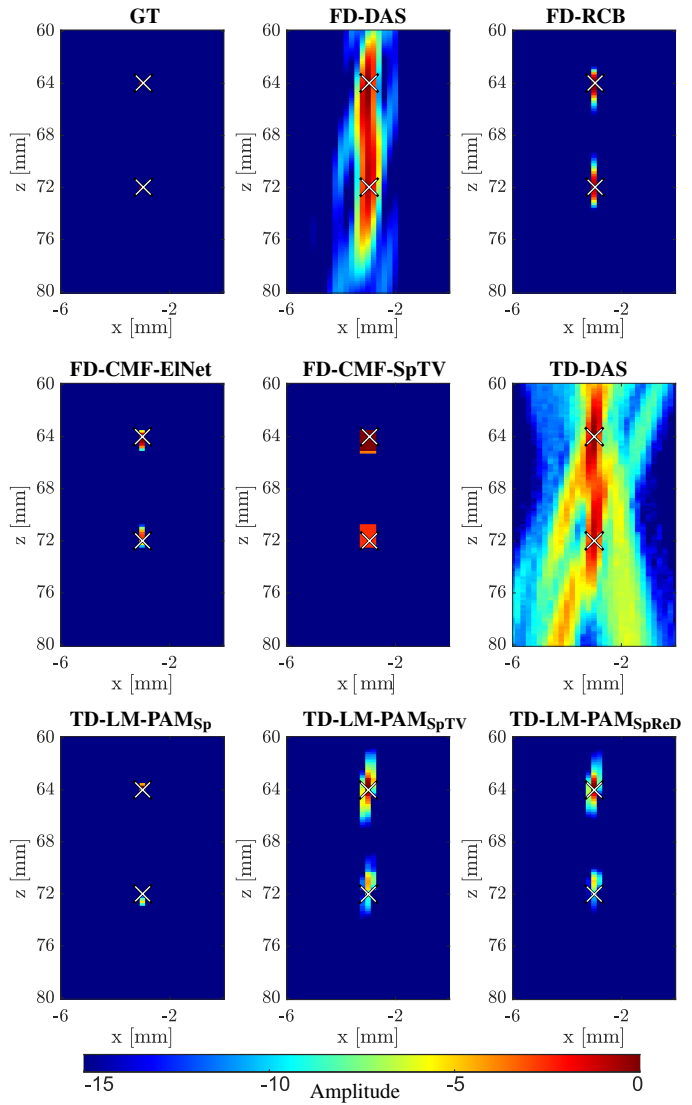

%% file: figures/figure5Arxiv.tex
\begin{figure}[b!]
    \centering
    \begin{tikzpicture}
                \node[anchor = north west, inner sep=0pt] (img) {\includegraphics[width=1\linewidth]{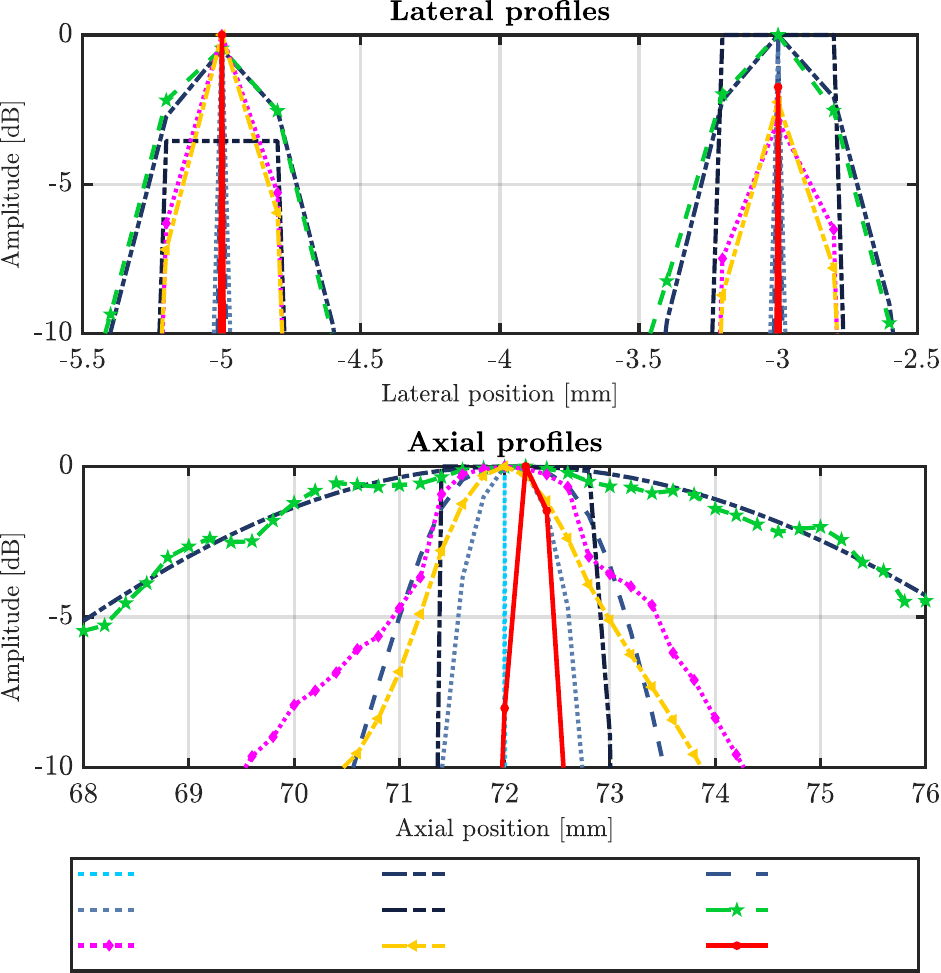}};
        \node[anchor  = north west, inner sep=0pt, align=center] 
        at (0.10\linewidth,-8.15) {\footnotesize{GT}};     
        \node[anchor  = north west, inner sep=0pt, align=center] 
        at (0.43\linewidth,-8.15) {\footnotesize{FD-DAS}};     
        \node[anchor  = north west, inner sep=0pt, align=center] 
        at (0.77\linewidth,-8.15) {\footnotesize{FD-RCB}};
        \node[anchor  = north west, inner sep=0pt, align=center] 
        at (0.10\linewidth,-8.53) {\footnotesize{FD-CMF-ElNet}};
        \node[anchor  = north west, inner sep=0pt, align=center] 
        at (0.43\linewidth,-8.53) {\footnotesize{FD-CMF-SpTV}};
         \node[anchor  = north west, inner sep=0pt, align=center] 
         at (0.77\linewidth,-8.53) {\footnotesize{TD-DAS}};
         \node[anchor  = north west, inner sep=0pt, align=center] 
         at (0.10\linewidth,-8.91) {\footnotesize{TD-LM-PAM$_{\text{SpTV}}$}};
        \node[anchor  = north west, inner sep=0pt, align=center] 
         at (0.43\linewidth,-8.91) {\footnotesize{TD-LM-PAM$_{\text{SpReD}}$}};
        \node[anchor  = north west, inner sep=0pt, align=center] 
         at (0.77\linewidth,-8.91) {\footnotesize{TD-LM-PAM$_{\text{Sp}}$}};
        \end{tikzpicture}
        \caption{Estimated axial and lateral profiles of two laterally distributed inertial microbubbles simulated at (-5, 72)~mm and (-3, 72)~mm across all evaluated methods.}
        \label{fig:ProfilesLateral}
    \end{figure}

%% file: figures/figure6Arxiv.tex
\begin{figure}[b!]
    \centering
    \begin{tikzpicture}
        \node[anchor = north west, inner sep=0pt] (img) {\includegraphics[width=1\linewidth]{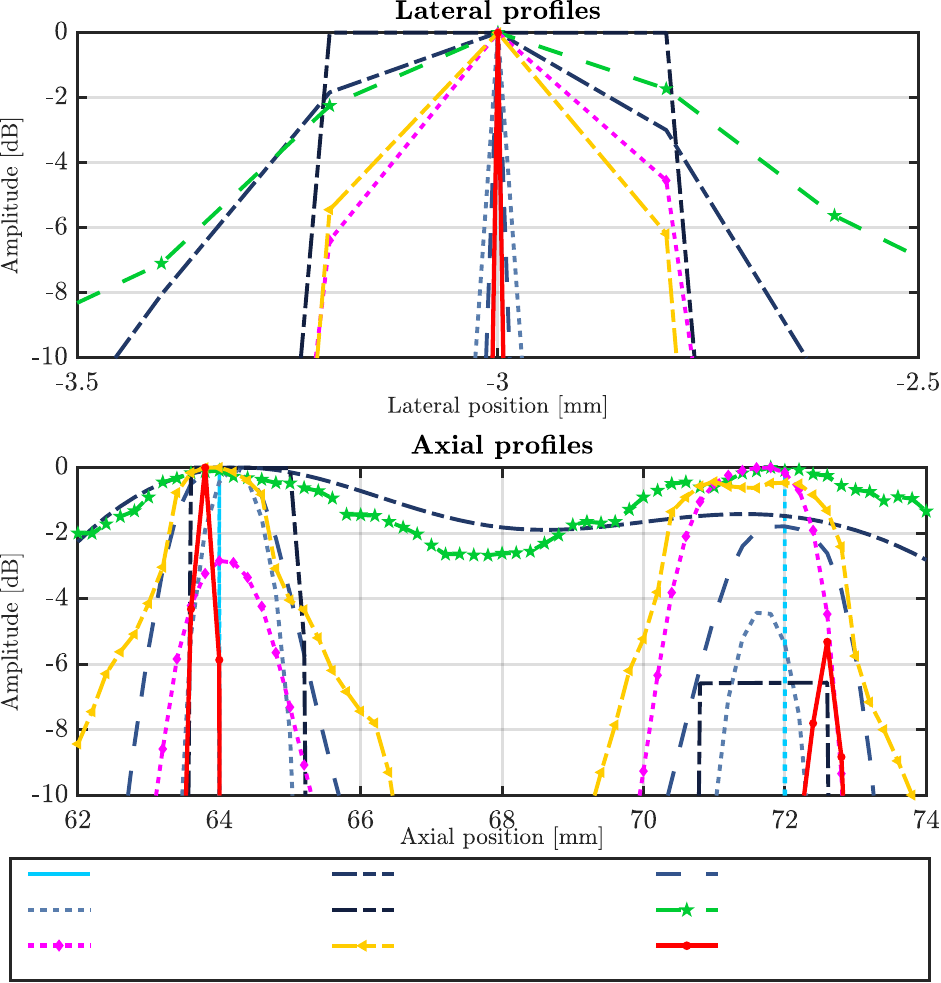}};
    \node[anchor  = north west, inner sep=0pt, align=center] 
        at (0.10\linewidth,-8.15) {\footnotesize{GT}};     
        \node[anchor  = north west, inner sep=0pt, align=center] 
        at (0.43\linewidth,-8.15) {\footnotesize{FD-DAS}};     
        \node[anchor  = north west, inner sep=0pt, align=center] 
        at (0.77\linewidth,-8.15) {\footnotesize{FD-RCB}};
        \node[anchor  = north west, inner sep=0pt, align=center] 
        at (0.10\linewidth,-8.53) {\footnotesize{FD-CMF-ElNet}};
        \node[anchor  = north west, inner sep=0pt, align=center] 
        at (0.43\linewidth,-8.53) {\footnotesize{FD-CMF-SpTV}};
         \node[anchor  = north west, inner sep=0pt, align=center] 
         at (0.77\linewidth,-8.53) {\footnotesize{TD-DAS}};
         \node[anchor  = north west, inner sep=0pt, align=center] 
         at (0.10\linewidth,-8.91) {\footnotesize{TD-LM-PAM$_{\text{SpTV}}$}};
        \node[anchor  = north west, inner sep=0pt, align=center] 
         at (0.43\linewidth,-8.91) {\footnotesize{TD-LM-PAM$_{\text{SpReD}}$}};
        \node[anchor  = north west, inner sep=0pt, align=center] 
         at (0.77\linewidth,-8.91) {\footnotesize{TD-LM-PAM$_{\text{Sp}}$}};
         \end{tikzpicture}
        \caption{Estimated axial and lateral profiles of two axially distributed inertial microbubbles simulated at (-3, 64)~mm and(-3, 72)~mm across all evaluated methods.}
        \label{fig:ProfilesAxial}
    \end{figure}

%% file: figures/figure7Arxiv.tex
\begin{figure}[b!]
    \centering
    \begin{tikzpicture}
        \node[anchor = north west, inner sep=0pt] (img) {\includegraphics[width=1\linewidth]{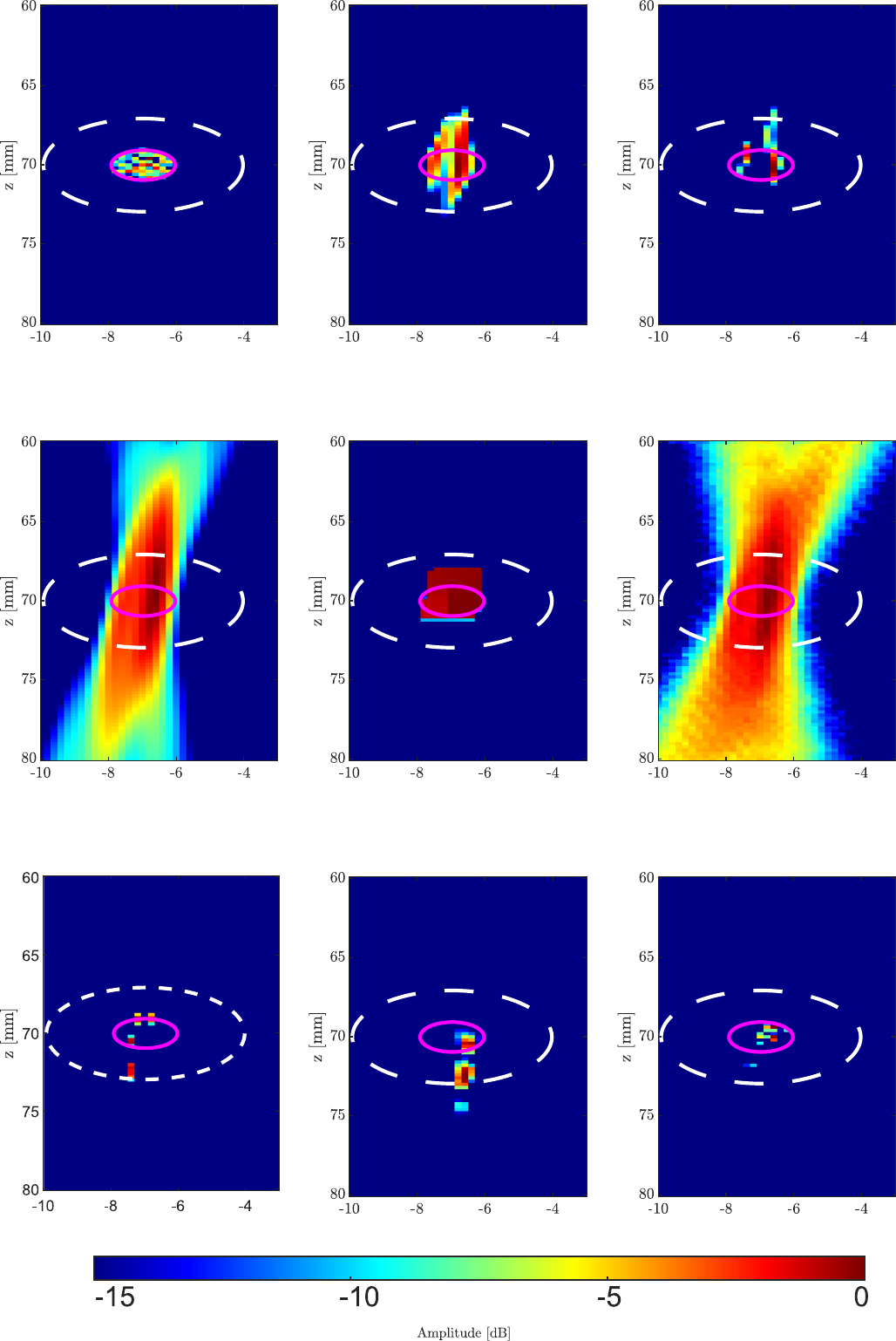}};
       \node[anchor  = north west, inner sep=0pt, align=center] 
        at (0.14\linewidth,0.3) {\footnotesize{\textbf{GT}}}; 
        \node[anchor  = north west, inner sep=0pt, align=center] 
        at (0.44\linewidth,0.3) {\footnotesize{\textbf{FD-DAS}}};
        \node[anchor  = north west, inner sep=0pt, align=center] 
        at (0.79\linewidth,0.3) {\footnotesize{\textbf{FD-RCB}}}; 
        \node[anchor  = north west, inner sep=0pt, align=center] 
        at (0.08\linewidth,-4.0) {\footnotesize{\textbf{FD-CMF-ElNet}}};
        \node[anchor  = north west, inner sep=0pt, align=center] 
        at (0.43\linewidth,-4.0) {\footnotesize{\textbf{FD-CMF-spTV}}};
        \node[anchor  = north west, inner sep=0pt, align=center] 
        at (0.8\linewidth,-4.0) {\footnotesize{\textbf{TD-DAS}}};
        \node[anchor  = north west, inner sep=0pt, align=center] 
        at (0.07\linewidth,-8.3) {\footnotesize{\textbf{TD-LM-PAM}$_{\text{\textbf{Sp}}}$}};
         \node[anchor  = north west, inner sep=0pt, align=center] 
        at (0.42\linewidth,-8.3) {\footnotesize{\textbf{TD-LM-PAM}$_{\text{\textbf{SpTV}}}$}};
         \node[anchor  = north west, inner sep=0pt, align=center] 
        at (0.74\linewidth,-8.3) {\footnotesize{\textbf{TD-LM-PAM}$_{\text{\textbf{SpReD}}}$}};
        \node[anchor  = north west, inner sep=0.03pt, align=center, fill = white] 
        at (0.45\linewidth,-13.0) {\footnotesize{Amplitude}}; 

\node[anchor  = north west, inner sep=0.03pt, align=center, fill = white, rotate = 90] 
        at (-0.01\linewidth,-2.45) {\footnotesize{Axial $z$~[mm]}}; 
\node[anchor  = north west, inner sep=0.03pt, align=center, fill = white, rotate = 90] 
        at (0.34\linewidth,-2.45) {\footnotesize{Axial $z$~[mm]}}; 
\node[anchor  = north west, inner sep=0.03pt, align=center, fill = white, rotate = 90] 
        at (0.68\linewidth,-2.45) {\footnotesize{Axial $z$~[mm]}};

        \node[anchor  = north west, inner sep=0.03pt, align=center, fill = white, rotate = 90] 
        at (-0.01\linewidth,-6.75) {\footnotesize{Axial $z$~[mm]}}; 
\node[anchor  = north west, inner sep=0.03pt, align=center, fill = white, rotate = 90] 
        at (0.34\linewidth,-6.75) {\footnotesize{Axial $z$~[mm]}}; 
\node[anchor  = north west, inner sep=0.03pt, align=center, fill = white, rotate = 90] 
        at (0.68\linewidth,-6.75) {\footnotesize{Axial $z$~[mm]}};

        \node[anchor  = north west, inner sep=0.03pt, align=center, fill = white, rotate = 90] 
        at (-0.01\linewidth,-11.05) {\footnotesize{Axial $z$~[mm]}}; 
\node[anchor  = north west, inner sep=0.03pt, align=center, fill = white, rotate = 90] 
        at (0.34\linewidth,-11.05) {\footnotesize{Axial $z$~[mm]}}; 
\node[anchor  = north west, inner sep=0.03pt, align=center, fill = white, rotate = 90] 
        at (0.68\linewidth,-11.05) {\footnotesize{Axial $z$~[mm]}};

      \node[anchor  = north west, inner sep=0.03pt, align=center, fill = white] 
        at (0.07\linewidth,-3.45) {\footnotesize{Lateral $x$~[mm]}}; 
        \node[anchor  = north west, inner sep=0.03pt, align=center, fill = white] 
        at (0.42\linewidth,-3.45) {\footnotesize{Lateral $x$~[mm]}}; 
           \node[anchor  = north west, inner sep=0.03pt, align=center, fill = white] 
        at (0.76\linewidth,-3.45) {\footnotesize{Lateral $x$~[mm]}};

      \node[anchor  = north west, inner sep=0.03pt, align=center, fill = white] 
        at (0.07\linewidth,-7.75) {\footnotesize{Lateral $x$~[mm]}}; 
        \node[anchor  = north west, inner sep=0.03pt, align=center, fill = white] 
        at (0.42\linewidth,-7.75) {\footnotesize{Lateral $x$~[mm]}}; 
           \node[anchor  = north west, inner sep=0.03pt, align=center, fill = white] 
        at (0.76\linewidth,-7.75) {\footnotesize{Lateral $x$~[mm]}}; 
        
        \node[anchor  = north west, inner sep=0.03pt, align=center, fill = white] 
        at (0.07\linewidth,-12.05) {\footnotesize{Lateral $x$~[mm]}}; 
        \node[anchor  = north west, inner sep=0.03pt, align=center, fill = white] 
        at (0.42\linewidth,-12.05) {\footnotesize{Lateral $x$~[mm]}}; 
           \node[anchor  = north west, inner sep=0.03pt, align=center, fill = white] 
        at (0.76\linewidth,-12.05) {\footnotesize{Lateral $x$~[mm]}}; 
        \end{tikzpicture}
        \caption{Estimated power maps of clouds. The white dotted zone corresponds to the noise area and the magenta zone corresponds to the signal area for CNR and Dice computation.}
        \label{fig:CloudMaps}
    \end{figure}
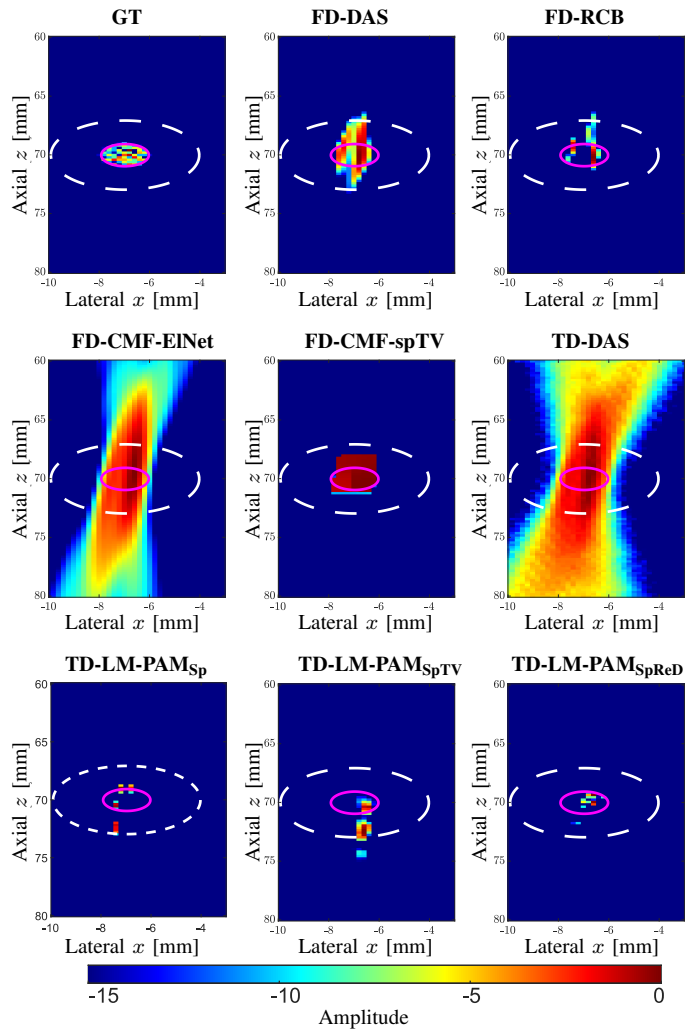

%% file: tables/table5.tex
\begin{table}[h!]
\centering
\small
\renewcommand{\arraystretch}{1.25}
\caption{Circular Cloud Setup Quantitative Performance}
\begin{tabularx}{\linewidth}{X|c|c}
\hline
\makecell[c]{\textbf{Method}} & \textbf{CNR {[}dB{]}} & \textbf{DICE {[}mm{]}} \\ \hline \hline
\textbf{FD-DAS} & 2.39 (0.08) & 0.49 (0.01) \\ \hline
\textbf{FD-RCB} & 3.06 (0.01) & 0.54 (0.02) \\ \hline
\textbf{FD-ENet} & 4 (4) & 0.2 (0.1) \\ \hline
\textbf{FD-SpTV} & 7 (4) & $\mathbf{0.8 \,(0.1)}$ \\ \hline
\textbf{TD-DAS} & 2.9 (0.4) & 0.35 (0.11) \\ \hline
\textbf{TD-LM-PAM$_\text{Sp}$} & 1.3 (2.6) & 0.25 (0.3) \\ \hline
\textbf{TD-LM-PAM$_{\text{SpTV}}$} & 3.8 (3.5) & 0.36 (0.29) \\ \hline
\textbf{TD-LM-PAM$_\text{SpRed}$} & $\mathbf{7.2(2.4)}$ & $\underline{0.76\,    (0.15)}$ \\ \hline
\end{tabularx}
\label{tab:CloudResults}
\end{table}